\documentclass[prb,onecolumn,superscriptaddress,showpacs]{revtex4}
\usepackage{psfrag,graphicx,stmaryrd,amsmath,amssymb}

\begin{document}


\title{Electronic Raman Scattering in Unconventional Density Waves}

\author{Andr\'as V\'anyolos}
\email{vanyolos@kapica.phy.bme.hu}
\affiliation{Department of Physics, Budapest University of Technology and Economics, 1521 Budapest, Hungary}
\author{Attila Virosztek}
\affiliation{Department of Physics, Budapest University of Technology and Economics, 1521 Budapest, Hungary}
\affiliation{Research Institute for Solid State Physics and Optics, PO Box 49, 1525 Budapest, Hungary.}

\begin{abstract}
We investigate the electronic Raman scattering in pure, quasi-one dimensional conductors with density wave ground state. In particular, we develop the theory of light-scattering on spin and charge density waves, both conventional and unconventional. We calculate the electronic Raman response of the interacting electron system with a single, highly anisotropic conduction band. The calculation is carried out in the mean field approximation. Beside the quasiparticle contribution, the electron-electron interaction is also included on RPA level. The contribution of collective modes and the effect of Coulomb screening are investigated. In analogy with unconventional superconductivity, the obtained Raman spectra - which are  finite in the low temperature phase possessing a gap, and vanish identically in the normal state - show unique and strong dependence on the polarization of the incoming and scattered light. We have found distinct, characteristic lineshapes, especially in the unconventional situation, depending on the various scattering geometries and the particular momentum dependence of the density wave order parameter.
\end{abstract}


\maketitle

\section{Introduction}

Electronic Raman scattering has been proven to be a valuable spectroscopic tool in identifying various low temperature phases of interacting electron systems. In early measurements on layered transition metal dichalcogenides like 2H-NbSe$_2$, a low energy peak was attributed to the charge density wave amplitude mode [\onlinecite{tsang}], while at even lower temperatures the superconducting gap also showed up in the Raman spectra [\onlinecite{sooryakumar}]. A series of papers investigated the Raman response of a system with competing spin density wave and superconducting instabilities [\onlinecite{Behera1,Behera2,Behera3,Behera4}]. Raman experiments contributed significantly to the establishment of the $d$-wave nature of the order parameter in high temperature superconductors (HTSC) [\onlinecite{leach}]. It has also been applied recently in order to investigate the temperature and pressure dependence of the charge density wave amplitude mode in 1$T$-TiSe$_2$ [\onlinecite{snow}]. Superconducting and density wave condensates, both unconventional, are believed to be present in the underdoped cuprates [\onlinecite{benfatto,bena,nayak}]. A theoretical analysis of this complex situation with respect to Raman scattering has also been attempted [\onlinecite{zeyher}]. Recent work on Raman spectra in HTSC calls attention to the importance of density fluctuations as well [\onlinecite{castro,tacon,caprara}].

The recent surge of interest in unconventional density waves (UDW) is mostly due to their potential applicability in the pseudogap phase [\onlinecite{janossy,opel,kaminski}] of HTSC materials. However pseudogap phases, and in general various kinds of hidden order, are detected in other substances as well, like in chalcogenides [\onlinecite{nemeth}], in heavy fermion materials [\onlinecite{isaacs}], and in Bechgaard salts [\onlinecite{christ}]. Since UDW's are natural candidates for explaining hidden order due to their momentum dependent gap structure [\onlinecite{balazs1}], they have been proposed to exist in URu$_2$Si$_2$ [\onlinecite{ikeda}] and in $\alpha$-(ET)$_2$ salts [\onlinecite{balazs2}]. Recent calculations of magnetoresistance, thermoelectric power and Nernst effect [\onlinecite{balazs4}] point out the possibility of UDW in (TMTSF)$_2$PF$_6$ [\onlinecite{balazs5}] and CeCoIn$_5$ [\onlinecite{balazs6}]. NMR results on Na$_{0.7}$CoO$_2$ are also consistent with the UDW scenario [\onlinecite{gavilano}].

The aim of the present paper is to develop a theory of Raman scattering in pure quasi-one dimensional conductors with conventional, or unconventional density wave ground state. Basics of electronic Raman scattering and mean field treatment of density waves are given in Section II. The quasiparticle contribution to the light scattering intensity in various polarizations and gap structures are calculated in Section III. In Section IV we incorporate the effect of electron-electron interaction. Namely we consider the collective contribution caused by the fluctuation of the order parameter in the usual RPA approximation, and also investigate the Coulomb screening. Finally Section V is devoted to our conclusions.

\section{Electronic Raman Scattering}

Light coupling to electrons via the vector potential $\mathbf{A}$ can be treated in second-order perturbation theory. The intensity of scattered light in a Raman experiment can be expressed [\onlinecite{dierker}] as
\begin{equation}
\frac{d\sigma}{d\omega d\Omega dV}=r_0^2\frac{\omega_s}{\omega_i}
S_{\gamma\gamma}(\mathbf{q},\omega),
\end{equation}
where $r_0^2=e^2/mc^2$ is the Thomson radius, $\omega_i,\mathbf{q}_i$ and $\omega_s,\mathbf{q}_s$ are the energies and momenta of the incoming and scattered photon, respectively. Furthermore the energy and momentum transfer to the material are $\omega=\omega_i-\omega_s$ and $\mathbf{q}=\mathbf{q}_i-\mathbf{q}_f$. The generalized structure factor $S_{\gamma\gamma}$ is related to the Raman response through the fluctuation-dissipation theorem
\begin{equation}\label{response}
S_{\gamma\gamma}(\mathbf{q},\omega)=\frac1\pi[1+n(\omega)]\text{Im}\chi_{\gamma\gamma}(\mathbf{q},\omega),
\end{equation}
where $n(\omega)$ is the Bose function. The Raman response of the electron system measures "effective density" fluctuations
\begin{equation}
\chi_{\gamma\gamma}(\mathbf{q},\omega)=i\langle[\tilde\rho(\mathbf{q}),\tilde\rho(\mathbf{-q})]\rangle(\omega)/V,
\end{equation}
where
\begin{equation}
\tilde\rho(\mathbf{q})=\sum_{\mathbf{k},\sigma}\gamma_\mathbf{k}c^\dag_{\mathbf{k+q},\sigma}c_{\mathbf{k},\sigma},
\end{equation}
$V$ is the volume of the system, and since we are interested in the $\mathbf{q}\to0$ behavior of the $\chi_{\gamma\gamma}$ susceptibility, we neglected the $\mathbf{q}$ dependence of the vertex $\gamma_\mathbf{k}$. Here $c^\dag_{\mathbf{k},\sigma},(c_{\mathbf{k},\sigma})$ is the creation (annihilation) operator of an electron with momentum $\mathbf{k}$ and spin $\sigma$ in the single conduction band $\epsilon_\mathbf{k}=-2t_a\cos ak_x -2t_b\cos bk_y -2t_c\cos ck_z$ with $t_a\gg t_b,t_c$. Our system is based on an orthorombic lattice with lattice constants $a,b,c$ towards the $x,y$ and $z$ directions. The strength of the scattering is determined by the momentum dependent function $\gamma_\mathbf{k}$ called the Raman vertex, which has the form
\begin{equation}
\gamma_\mathbf{k}=(\mathbf{e}_i\mathbf{e}_s)
+\frac{1}{m}\sum_b \left(\frac{\langle\mathbf{k}|\mathbf{pe}_s|b\mathbf{k}\rangle
\langle b\mathbf{k}|\mathbf{pe}_i|\mathbf{k}\rangle}{\epsilon_\mathbf{k}-\epsilon_{b\mathbf{k}}+\omega_i}+
\frac{\langle\mathbf{k}|\mathbf{pe}_i|b\mathbf{k}\rangle
\langle b\mathbf{k}|\mathbf{pe}_s|\mathbf{k}\rangle}{\epsilon_\mathbf{k}-\epsilon_{b\mathbf{k}}-\omega_s}\right),
\end{equation}
where $b$ stands for the band index of the electron excited out of the conduction band, and the corresponding states are $|\mathbf{k}b\rangle$ and $|\mathbf{k}\rangle$, respectively. In addition the polarization vectors of the incoming and scattered light are denoted by $\mathbf{e}_i,\mathbf{e}_s$. If the incoming and scattered light frequencies can be neglected in comparison to the optical band gap [\onlinecite{genkin}], the Raman vertex is related to the inverse mass tensor $\gamma_{\alpha\beta}(\mathbf{k})=m\partial^2\epsilon_\mathbf{k}/\partial k_\alpha \partial k_\beta$ through the relation
\begin{equation}\label{gamma}
\gamma_\mathbf{k}=\sum_{\alpha,\beta}e_s^\alpha\gamma_{\alpha\beta}(\mathbf{k})e_i^\beta,
\end{equation}
which is widely known as the effective mass approximation.

The retarded susceptibility of the effective density can be evaluated with analytical continuation from the Fourier transform of the corresponding $\tau$ (imaginary time) ordered response $\chi_{\gamma\gamma}(\mathbf{q},\tau)=-\langle T_\tau[\tilde\rho(\mathbf{q},\tau)\tilde\rho(-\mathbf{q})]\rangle$ in the usual way. The one particle Green's function of the DW using Nambu's notation reads
\begin{equation}
G(\mathbf{k},i\omega_n)=-\int_{0}^{\beta}\text{d}\tau\,\langle T_\tau[\Psi(\mathbf{k},\tau)
\Psi^\dag(\mathbf{k})]\rangle e^{i\omega_n\tau},
\end{equation}
where the four component spinor field
\begin{equation}\label{spinor}
\Psi(\mathbf{k},\tau)=
\begin{pmatrix}
c_{\mathbf{k},\shortuparrow}(\tau)\\
c_{\mathbf{k-Q},\shortuparrow}(\tau)\\
c_{\mathbf{k},\shortdownarrow}(\tau)\\
c_{\mathbf{k-Q},\shortdownarrow}(\tau)
\end{pmatrix}
\end{equation}
is introduced to simultaneously cover the spin space and to treat the left- and right-moving electrons in momentum space in a convenient way. $\mathbf{Q}=(2k_F,\pi/b,\pi/c)$ is the best nesting vector. Now the Green's function of an USDW is obtained as
\begin{equation}\label{green}
G^{-1}(\mathbf{k},i\omega_n)=i\omega_n-\xi_\mathbf{k} \rho_3-\Delta(\mathbf{k})\rho_1\sigma_3,
\end{equation}
while for UCDW $\sigma_3$ has to be replaced by one. We note at this point, that we assumed a real order parameter $\Delta(\mathbf{k})$, as in the absence of impurities and pinning the phase is unrestricted and therefore can be chosen to be zero for convenience. Here $\rho_i$, $(\sigma_i)$ are the Pauli matrices acting on momentum (spin) space respectively, while the linearized spectrum of the highly anisotropic electron system around the Fermi energy is $\xi_\mathbf{k}=\epsilon_\mathbf{k}-\mu=v_F(k_x-k_F)-2t_b\cos(bk_y)-2t_c\cos(ck_z)$. The order parameter is either independent of the momentum, which is the case of a conventional DW, or it can have four different type of wavevector dependence ($\Delta(\mathbf{k})=\Delta\cos bk_y$, $\Delta(\mathbf{k})=\Delta\sin bk_y$, $\Delta(\mathbf{k})=\Delta\cos ck_z$, $\Delta(\mathbf{k})=\Delta\sin ck_z$) as discussed in detail in  Ref.~[\onlinecite{balazs1}]. Henceforth, without the loss of generality we can assume a $k_y$ dependent gap to be open, since the two perpendicular directions are equivalent in our model.

\section{Quasiparticle contribution}

Making use of the anisotropic nearest neighbor tight-binding band structure and Eq.~\eqref{gamma}, the Raman tensor becomes diagonal with the same cosine functions being in the diagonal that appear in $\epsilon_\mathbf{k}$. Since the band structure belongs to the completely symmetric irreducible representation $A_g$ of the pointgroup of the lattice $D_{2h}$, it follows that this is similarly true for every component of the vertex $\gamma_\mathbf{k}$. Our model is therefore only capable of describing the Raman spectra belonging to the $A_g$ symmetry channel, \emph{i.e.} the spectra measured in $x-x$, $y-y$ and $z-z$ scattering geometries. In order to generalize the present treatment to incorporate the possibility of scattering effects with perpendicular polarizations, for instance the $x-y$ geometry, nonvanishing offdiagonal components of the Raman tensor are needed. Particularly in the $x-y$ geometry, the inclusion of a second nearest neighbor hopping term in the $a-b$ plane in the one particle energy can account for finite absorption. It can be readily shown that in an orthorombic lattice the spectrum obtained in the $x-y$ geometry belongs to the $B_{1g}$ representation. We return to this point at the end of this section.

\subsection{Raman spectra with $A_g$ symmetry}

Coming back to the Green's function in Eq.~\eqref{green}, the quasiparticle contribution to the Raman susceptibility for (U)DW can be written as the sum of three terms corresponding to the three different polarization directions
\begin{equation}
\begin{split}
\chi_{\gamma\gamma}(\mathbf{q},i\omega_n)&=
-\frac{1}{\beta V}\sum_{\mathbf{k},\omega_m}\text{Tr}(
\Gamma(\mathbf{k})G(\mathbf{k},i\omega_m)
\Gamma(\mathbf{k-q})G(\mathbf{k-q},i\omega_m-i\omega_n))\\
&=\chi_{\gamma\gamma}^x+\chi_{\gamma\gamma}^y+\chi_{\gamma\gamma}^z,
\end{split}
\end{equation}
where $\Gamma(\mathbf{k})$ is a four by four diagonal matrix with the elements $(\gamma_\mathbf{k},\gamma_\mathbf{k-Q},\gamma_\mathbf{k},\gamma_\mathbf{k-Q})$ in the diagonal. For simplicity, we shall limit our analysis to $\mathbf{q}=(q_x,0,0)$ (\emph{i.e.} wave vector pointing in the quasi-one-dimensional direction). For the retarded correlation functions we get
\begin{subequations}\label{raman}
\begin{eqnarray}
\chi_{\gamma\gamma}^x(\xi,\omega) &=&
2g(0)\gamma_x^2\left\{
\mu^2\frac{\xi^2}{\xi^2-\omega^2}(1-4\Delta^2F_2)+
\Delta^2\left(\frac{2}{\rho(0)P}+\omega^2F_2-\frac{4\Delta^2\omega^4}{(\omega^2-\xi^2)^2}F_4\right)
\right\},\label{raman-x}\\
\chi_{\gamma\gamma,\text{sin}}^y(\xi,\omega) &=&
4g(0)\gamma_y^2t_b^2\frac{\xi^2}{\xi^2-\omega^2}\left(1-\frac{8\omega^2\Delta^2}{\xi^2}
(F_2-F_4)\right),\label{raman-y-sin}\\
\chi_{\gamma\gamma,\text{cos}}^y(\xi,\omega) &=&
4g(0)\gamma_y^2t_b^2\frac{\xi^2}{\xi^2-\omega^2}\left(1-\frac{8\omega^2\Delta^2}{\xi^2}
F_4\right),\label{raman-y-cos}\\
\chi_{\gamma\gamma}^z(\xi,\omega) &=&
4g(0)\gamma_z^2t_c^2\frac{\xi^2}{\xi^2-\omega^2}\left(1-\frac{4\omega^2\Delta^2}
{\xi^2}F_2\right),\label{raman-z}
\end{eqnarray}
\end{subequations}
where $g(0)$, ($\rho(0)$) is the density of states at the Fermi energy in the normal state per spin per unit volume (per unit cell), $\xi=v_Fq_x$, $\mu$ is the chemical potential and $\gamma_x=ma^2(e^i_x e^s_x)$, $\gamma_y=mb^2(e^i_y e^s_y)$, $\gamma_z=mc^2(e^i_z e^s_z)$. Furthermore $P$ is the relevant coupling responsible for the DW formation whose detailed form can be found in Ref.~[\onlinecite{balazs1}], while
\begin{eqnarray}\label{f-function}
F_n&=&(\xi^2-\omega^2)\frac{1}{2\pi}\int_0^\infty\int_0^{2\pi}\tanh\left(\frac{\beta E}{2}\right)\frac{N}{D}
\text{Re}\frac{\sin^n(y)}{\sqrt{E^2-\Delta^2\sin^2(y)}}\,\text{d}y\,\text{d}E\\
N&=&(\xi^2-\omega^2)^2-4E^2(\xi^2+\omega^2)+4\Delta^2\xi^2\sin^2(y)\notag\\
D&=&N^2-64E^2\omega^2\xi^2(E^2-\Delta^2\sin^2(y))\notag
\end{eqnarray}
is the $F$ function that shows up in the correlation functions of conventional DWs with constant gap [\onlinecite{viro2}], as well as in the unconventional DWs [\onlinecite{balazs3}]. Eqs.~\eqref{raman} correspond to single bubble diagrams with self energy corrections due to the order parameter of the condensate.

\begin{figure}
\psfrag{x1}[t][b]{$\omega/2\Delta$} 
\psfrag{y1}[b][t]{$\text{Im}\chi_{\gamma\gamma}^x(\omega)/2g(0)\gamma_x^2\Delta^2$}
\psfrag{z1}[b][b]{}
\hfill
\rotatebox{-90}{\includegraphics[width=5.7cm]{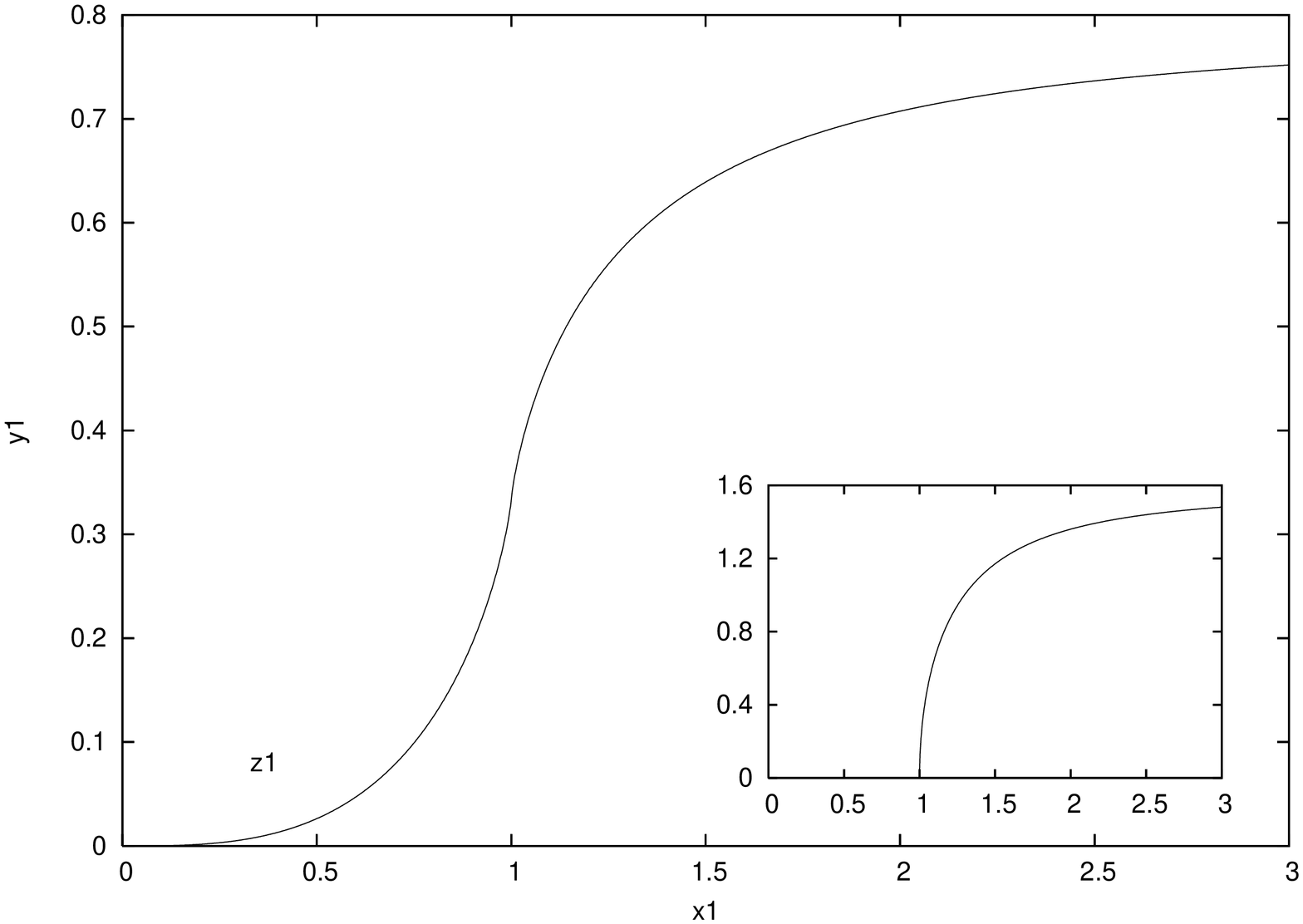}}\hfill
\psfrag{x4}[t][b]{$\omega/2\Delta$}
\psfrag{y4}[b][t]{$\text{Im}\chi_{\gamma\gamma}^z(\omega)/2g(0)\gamma_z^2 t_c^2$}
\psfrag{z4}[br][br]{}
\rotatebox{-90}{\includegraphics[width=5.7cm]{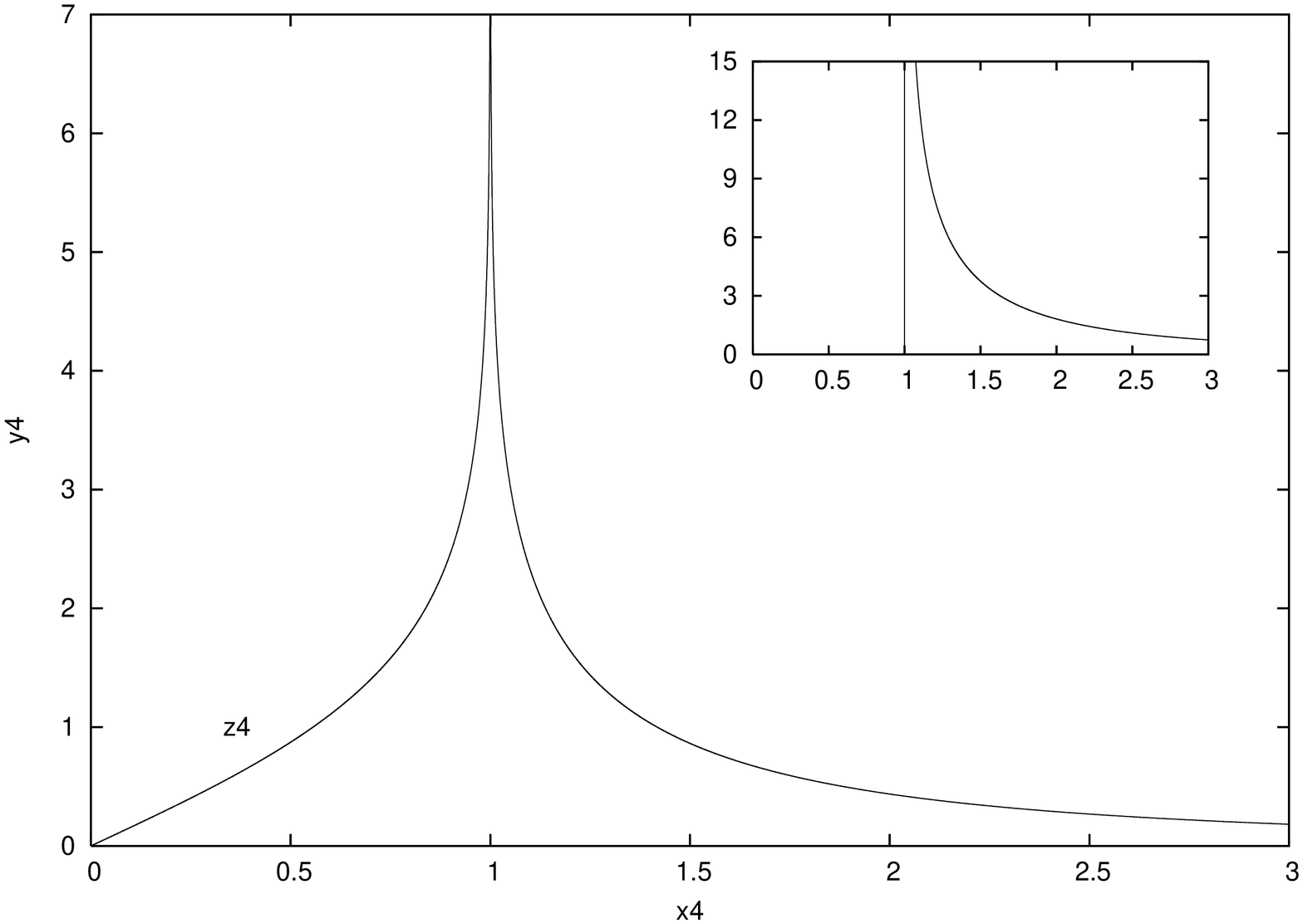}}
\caption{\label{fig:xz} Raman spectrum of an UDW for $x-x$ polarization (left panel), $z-z$ polarization (right panel) at $T=0$. Insets: the same spectra in a conventional DW.}
\end{figure}

\begin{figure}
\psfrag{x2}[t][b]{$\omega/2\Delta$}
\psfrag{y2}[b][t]{$\text{Im}\chi_{\gamma\gamma,\text{sin}}^y(\omega)/2g(0)\gamma_y^2 t_b^2$}
\psfrag{z2}[br][br]{}
\hfill
\rotatebox{-90}{\includegraphics[width=5.7cm]{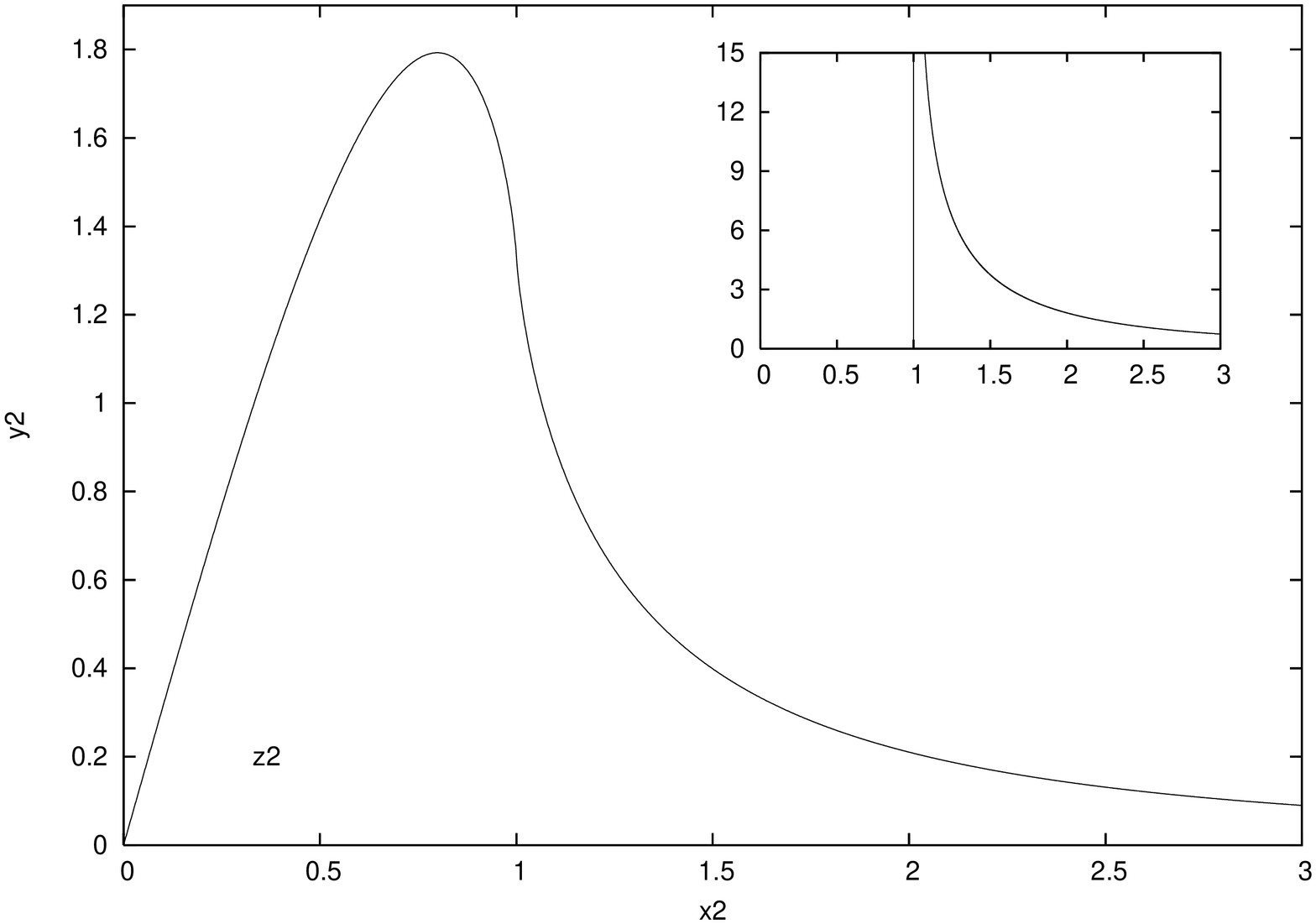}}\hfill
\psfrag{x3}[t][b]{$\omega/2\Delta$}
\psfrag{y3}[b][t]{$\text{Im}\chi_{\gamma\gamma,\text{cos}}^y(\omega)/2g(0)\gamma_y^2 t_b^2$}
\psfrag{z3}[b][b]{}
\rotatebox{-90}{\includegraphics[width=5.7cm]{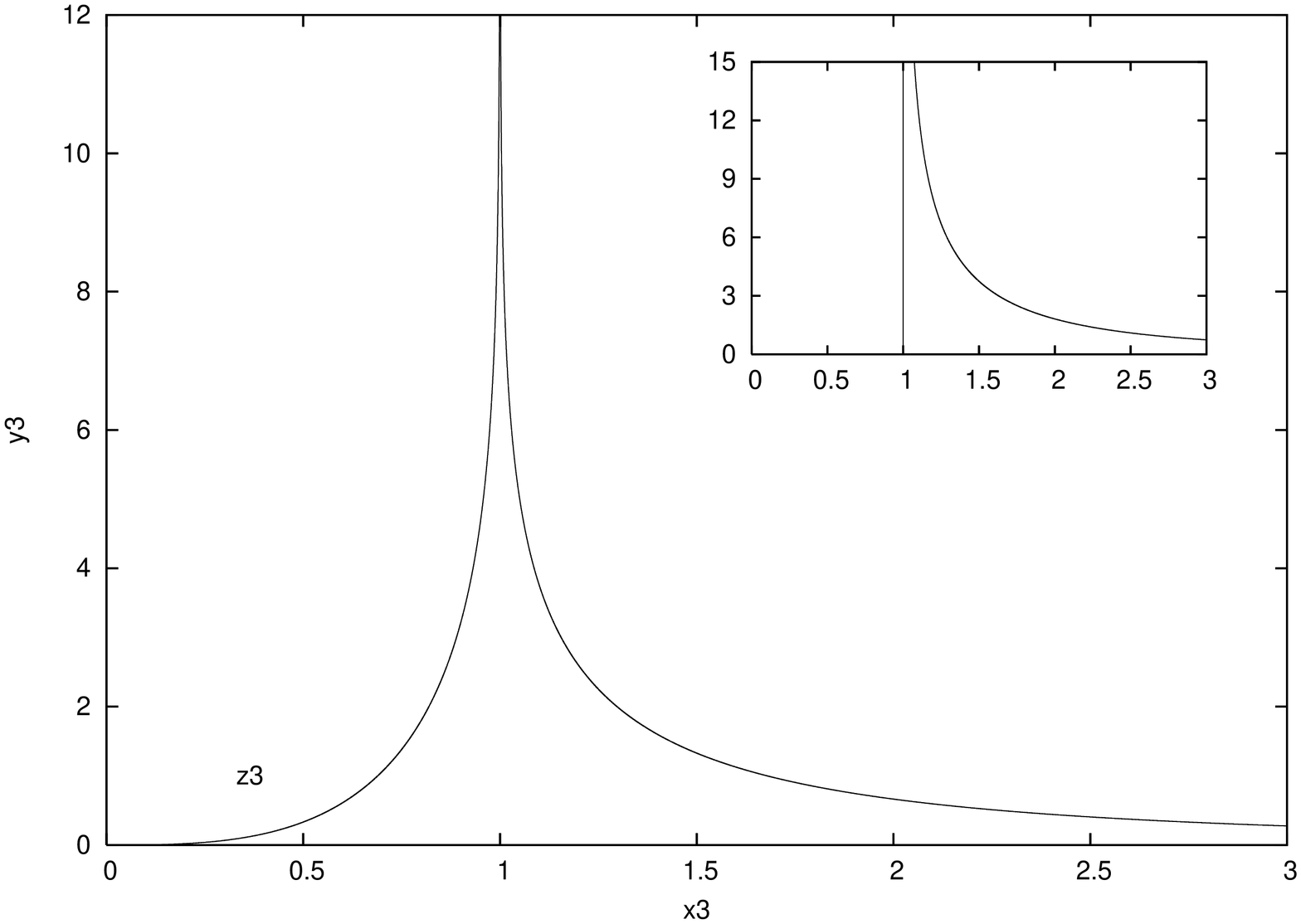}}
\caption{\label{fig:y} Raman spectrum of an UDW for $y-y$ polarization with $\Delta(\mathbf{k})=\Delta\sin(bk_y)$ (left panel), with $\Delta(\mathbf{k})=\Delta\cos(bk_y)$ (right panel). Insets: the same spectra in a conventional DW.}
\end{figure}

Since the momentum transfer of scattered light is small compared to the Fermi wave vector, we are only interested in the $\xi\to 0$ long wavelength limit. Taking the imaginary part of the obtained susceptibilities according to Eq.~\eqref{response}, at zero temperature we find for conventional DW
\begin{subequations}\label{conv}
\begin{eqnarray}
\text{Im}\chi^x_{\gamma\gamma,\text{conv}}&=&2g(0)\gamma_x^2\Delta^2\frac{\pi}{2x}\text{Re}\sqrt{x^2-1},\\
\text{Im}\chi^{y,z}_{\gamma\gamma,\text{conv}}&=&2g(0)\gamma_{y,z}^2t_{b,c}^2\frac{2\pi}{x}\text{Re}\frac{1}{\sqrt{x^2-1}},
\end{eqnarray}
\end{subequations}
while for unconventional DW we get
\begin{subequations}\label{elliptic}
\begin{eqnarray}
\text{Im}\chi^x_{\gamma\gamma}&=&\frac{2g(0)\gamma_x^2\Delta^2}{3x}
\left\{
\begin{array}{ll}
2(x^2-1)K(x)-(x^2-2)E(x), & x<1,\\
x(x^2-1)K(1/x)-x(x^2-2)E(1/x), & x\ge 1,
\end{array}
\right.\label{elliptic-x}\\
\text{Im}\chi^y_{\gamma\gamma,\text{sin}}&=&\frac{8g(0)\gamma_y^2t_b^2}{3x}
\left\{
\begin{array}{ll}
(1-x^2)K(x)-(1-2x^2)E(x), & x<1,\\
2x(1-x^2)K(1/x)-x(1-2x^2)E(1/x), & x\ge 1,
\end{array}
\right.\label{elliptic-y-sin}\\
\text{Im}\chi^y_{\gamma\gamma,\text{cos}}&=&\frac{8g(0)\gamma_y^2t_b^2}{3x}
\left\{
\begin{array}{ll}
(x^2+2)K(x)-2(x^2+1)E(x), & x<1,\\
x(2x^2+1)K(1/x)-2x(x^2+1)E(1/x), & x\ge 1,
\end{array}
\right.\label{elliptic-y-cos}\\
\text{Im}\chi^z_{\gamma\gamma}&=&\frac{4g(0)\gamma_z^2t_c^2}{x}
\left\{
\begin{array}{ll}
K(x)-E(x), & x<1,\\
x(K(1/x)-E(1/x)), & x\ge 1,
\end{array}
\right.\label{elliptic-z}
\end{eqnarray}
\end{subequations}
where $x=\omega/2\Delta$ and $K(x)$, $E(x)$ are the complete elliptic integrals of the first and second kind, respectively. The response functions for finite $T$ are obtained simply by multiplying Eqs.~\eqref{conv} and \eqref{elliptic} by $\tanh(\omega/4T)$. The spectra are shown in Figs.~\ref{fig:xz} and \ref{fig:y}.

For conventional DW - due to the equivalence of the $b$ and $c$ crystal directions - not surprisingly we obtain the same response for the $y-y$ and $z-z$ polarizations showing the usual inverse squareroot divergence at $2\Delta$. It is worth mentioning, that for the chain polarization $x-x$ the divergent peak is suppressed and transformed to a squareroot edge at the same threshold due to the vanishing vertex on the Fermi surface.
It turns out that the vertex responsible for the scattering intensity is $(\gamma_\mathbf{k}-\gamma_\mathbf{k-Q})^2$ in our Nambu notation, which in this particular scattering geometry is proportional to $\xi_\mathbf{k}^2$ that clearly vanishes on the Fermi surface.

In contrast to conventional DW, in an UDW there are linenodes on the Fermi surface [\onlinecite{balazs1}], giving rise to arbitrarily small energy nodal-excitations. It follows that the scattering intensity is finite for frequencies smaller than the maximum optical gap $2\Delta$. Furthermore, in the $y-y$ geometry the interplay between the cosinusoidal vertex and the order parameter - either sinusoidal or cosinusoidal - results in two qualitatively different lineshapes (see Fig.~\ref{fig:y}). The clear singularities in the spectra at $2\Delta$ for UDW that appear in Figs.~\ref{fig:xz} and \ref{fig:y} are of logarithmic type and are caused by Van Hove singularities in the quasiparticle density of states [\onlinecite{balazs1}]. However this peak in Fig.~\ref{fig:y} (left panel) is suppressed because of the zero Raman vertex at the gap maximum.

The low frequency power law behavior is also characteristic for systems with point- or linenodes on the Fermi surface. In particular we have
\begin{subequations}\label{lowfreq}
\begin{eqnarray}
\text{Im}\chi^x_{\gamma\gamma}(\omega\to0)&=&2g(0)\gamma_x^2\Delta^2\frac{\pi}{16}\left(\frac{\omega}{2\Delta}\right)^3 + \mathcal{O}(\omega^5),\\
\text{Im}\chi^y_{\gamma\gamma,\text{sin}}(\omega\to0)&=&2g(0)\gamma_y^2t_b^2\pi\frac{\omega}{2\Delta} + \mathcal{O}(\omega^3),\\
\text{Im}\chi^y_{\gamma\gamma,\text{cos}}(\omega\to0)&=&2g(0)\gamma_y^2t_b^2\frac{3\pi}{4}\left(\frac{\omega}{2\Delta}\right)^3 + \mathcal{O}(\omega^5),\\
\text{Im}\chi^z_{\gamma\gamma}(\omega\to0)&=&2g(0)\gamma_z^2t_c^2\frac{\pi}{2}\frac{\omega}{2\Delta} + \mathcal{O}(\omega^3).
\end{eqnarray}
\end{subequations}
All these important features of the Raman response makes the Raman experiment to be a relevant and fruitful probe in
identifying the magnitude and symmetry of the energy gap. Similar analysis [\onlinecite{devereaux-1}] contributed to the
establishment of the $d$-wave nature of the order parameter in HTSC.

\subsection{Raman spectra belonging to $B_{1g}$, $B_{2g}$ and $B_{3g}$ symmetries}

We have already pointed out in the beginning of this section that the choice of the nearest neighbor tight-binding band structure for the one particle energies is only sufficient to describe the Raman response in the $A_g$ symmetry channel. Now we extend the previous analysis with the inclusion of second nearest neighbor hopping terms in the $a-b$, $a-c$ and $b-c$ crystal planes, respectively. This extension on one hand can be considered as the simplest natural and physically motivated step towards the treatment of more realistic one particle energies, on the other hand it is sufficient to explain the Raman spectra measured with perpendicular polarizations that belong to the other three irreducible representation of the pointgroup. Namely, we add to $\epsilon_\mathbf{k}$ the following extra term
\begin{equation}
\delta\epsilon_\mathbf{k}=4t_{xy}\cos(ak_x)\cos(bk_y)+4t_{xz}\cos(ak_x)\cos(ck_z)+4t_{yz}\cos(bk_y)\cos(ck_z).\label{extra}
\end{equation}
The first term will clearly give nonzero offdiagonal component in the Raman tensor $(\gamma^{xy}\sim\sin(ak_x)\sin(bk_y))$. This function of the wavevector $\mathbf{k}$ is the most simple basis function belonging to the $B_{1g}$ representation of $D_{2h}$. Similarly the second and third terms of Eq.~\eqref{extra} yield nonvanishing contributions in the other two offdiagonal positions of the tensor, and the corresponding functions belong to the remaining two representations $B_{2g}$ and $B_{3g}$, respectively.

Now making use of the formalism and notations introduced in the previous subsection, the Raman spectra in the $x-y$ scattering geometry - labelled by the $B_{1g}$ symmetry - are obtained as
\begin{subequations}\label{b1g-ch}
\begin{eqnarray}
\chi_{\gamma\gamma,\text{sin}}^{B_{1g}}(\xi,\omega) &=& 4g(0)\gamma_{xy}^2\left\{
\alpha t_{xy}^2\frac{\xi^2}{\xi^2-\omega^2}(1-8\Delta^2F_4)+
2\beta\Delta^2\left(\frac{1}{\rho(0)P}+\omega^2F_4-\frac{4\Delta^2\omega^4}{(\omega^2-\xi^2)^2}
F_6\right)
\right\},\label{b1g-sin}\\
\chi_{\gamma\gamma,\text{cos}}^{B_{1g}}(\xi,\omega) &=& 4g(0)\gamma_{xy}^2\left\{
\alpha t_{xy}^2\frac{\xi^2}{\xi^2-\omega^2}(1-8\Delta^2(F_2-F_4))\right.\notag\\
&\phantom{=}& +2\beta\Delta^2\left.\left(\frac{1}{\rho(0)P}+\omega^2(F_2-F_4)-\frac{4\Delta^2\omega^4}{(\omega^2-\xi^2)^2}
(F_4-F_6)\right)
\right\}.\label{b1g-cos}
\end{eqnarray}
\end{subequations}
In the $x-z$ geometry we have
\begin{eqnarray}\label{b2g}
\chi_{\gamma\gamma}^{B_{2g}}(\xi,\omega) &=& 4g(0)\gamma_{xz}^2\left\{
\alpha t_{xz}^2\frac{\xi^2}{\xi^2-\omega^2}(1-4\Delta^2F_2)+
\beta'\Delta^2\left(\frac{2}{\rho(0)P}+\omega^2F_2-\frac{4\Delta^2\omega^4}{(\omega^2-\xi^2)^2}
F_4\right)\right\},
\end{eqnarray}
and finally in the $y-z$ geometry we get
\begin{subequations}\label{b3g-ch}
\begin{eqnarray}
\chi_{\gamma\gamma,\text{sin}}^{B_{3g}}(\xi,\omega) &=&
8g(0)\gamma_{yz}^2t_{yz}^2\frac{\xi^2}{\xi^2-\omega^2}\left(1-8\Delta^2 F_4\right),\label{b3g-sin}\\
\chi_{\gamma\gamma,\text{cos}}^{B_{3g}}(\xi,\omega) &=&
8g(0)\gamma_{yz}^2t_{yz}^2\frac{\xi^2}{\xi^2-\omega^2}\left(1-8\Delta^2(F_2-F_4)\right).\label{b3g-cos}
\end{eqnarray}
\end{subequations}
Here $F_n$ is given by Eq.~\eqref{f-function}, the superscripts \emph{sin} and \emph{cos} in Eqs.~\eqref{b1g-ch} and \eqref{b3g-ch} denote the wavevector dependence of the order parameter $\Delta(\mathbf{k})$ as in Eqs.~\eqref{raman}. Furthermore $\alpha=4\sin^2(ak_F)$, $\beta=(t_{xy}/t_a)^2\cot^2(ak_F)$, $\beta'=(t_{xz}/t_a)^2\cot^2(ak_F)$, $\gamma_{xy}=mab$, $\gamma_{xz}=mac$, and $\gamma_{yz}=mbc$. In the long wavelength limit $\xi\to0$ and $\chi^{B_{3g}}$ clearly vanishes, while $\chi^{B_{1g}}$ and $\chi^{B_{2g}}$ remain finite. The spectra emanating from the latter two are qualitativly not different from $\text{Im}\chi_{\gamma\gamma}^x$ plotted on Fig.~\ref{fig:xz} left panel, therefore are not shown here. Nevertheless, the above considerations indicate, that assuming a more realistic band structure would not significantly alter the Raman lineshapes obtained in this section.

\section{Electron-electron interaction, RPA series}

The previous section dealt with the Raman response function in the one bubble approximation, i.e. the effect of interaction is taken into account in the self energy only. Now we turn our attention to vertex corrections at the RPA level, since the short and long wavelength components of the interaction may give rise to collective modes and Coulomb screening, respectively.

\subsection{Collective excitations in the Raman response}

Following Ref.~[\onlinecite{balazs3}], the short wavelength component of the electron-electron interaction favoring a sinusoidal gap in the $k_y$ direction, namely $\Delta(\mathbf{k})=\Delta\sin(bk_y)$, is given by
\begin{equation}\label{int-matrix}
\begin{split}
\frac{N}{V}\tilde V(\mathbf{k,k',q},\sigma,\sigma')&=\delta_{-\sigma,\sigma'}
\left(2J_y\sin(bk_y)\sin(b(k'_y-q_y))-2F_y\sin(bk_y)\sin(bk'_y)\right)\\
&\quad+\delta_{\sigma,\sigma'}(J_y-V_y)\sin(bk_y)\sin(b(k'_y-q_y)).
\end{split}
\end{equation}
The detailed form of the whole interaction responsible for the density wave formation with the relevant couplings ($P$, see Eq.~\eqref{int}) can be found in Ref.~[\onlinecite{balazs1}]. Here we shall continue with the assumption we made in Section II, namely that as we enter the low temperature phase (LTP), a gap varying in the $k_y$ direction opens up first and persists all the way down to zero temperature. Moreover we fix its functional form to be sinusoidal. All the calculations we present here can also be done with cosinusoidal gap without any relevant changes. 

\begin{figure}
\psfrag{x}[t][b]{$\omega/2\Delta$}
\psfrag{y}[b][t]{$-(\omega^2F_2-4\Delta^2F_4)'$ and $(\omega^2F_2-4\Delta^2F_4)''$}
\hfill
\rotatebox{-90}{\includegraphics[width=6.5cm,height=8.1cm]{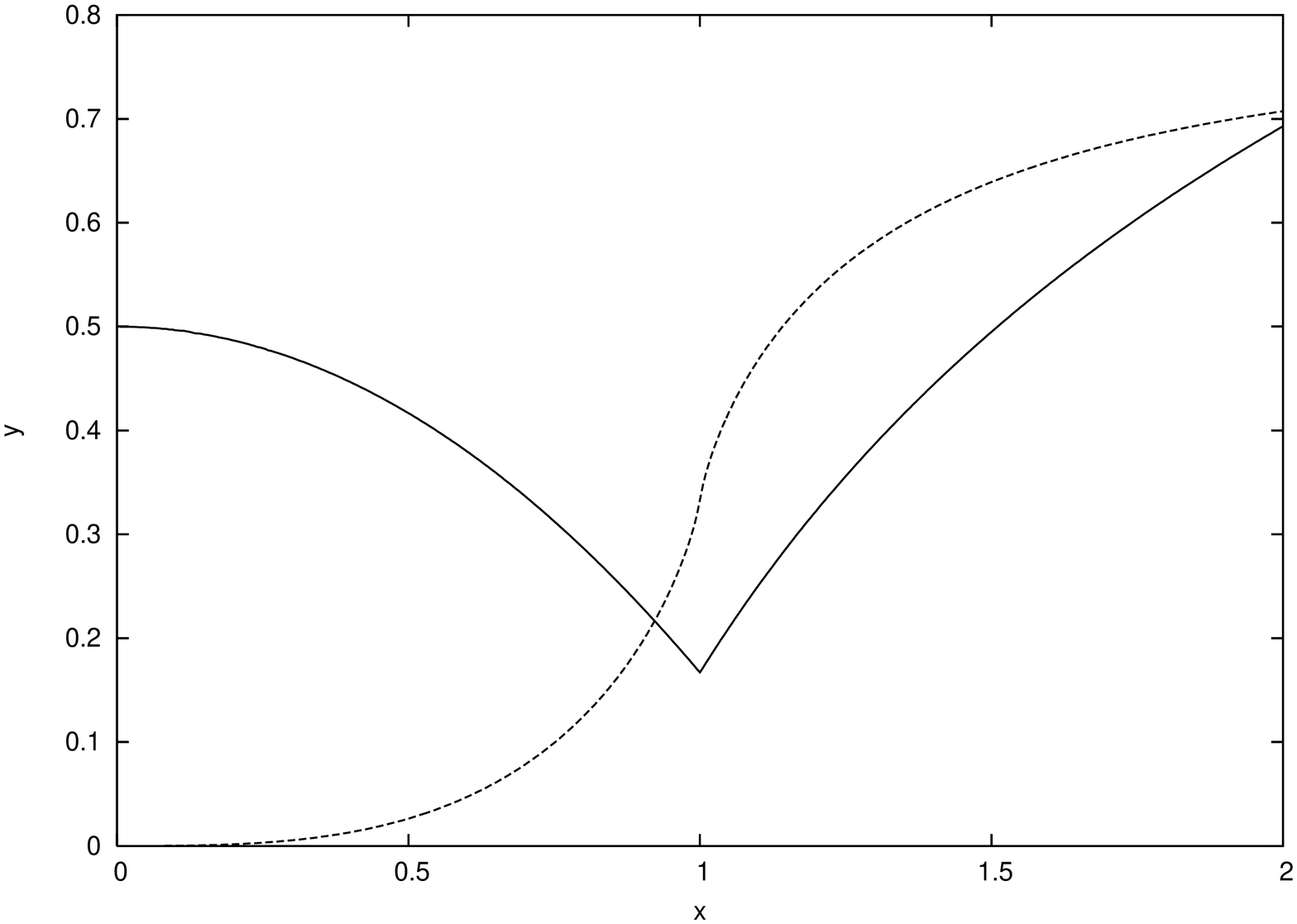}}\hfill
\psfrag{x}[t][b]{$\omega/2\Delta$}
\psfrag{y}[b][t]{$\text{Im}\chi_{\gamma\gamma}^x(\omega)/2g(0)\gamma_x^2(\Delta/\lambda)^2$}
\psfrag{z}{}
\rotatebox{-90}{\includegraphics[width=6.5cm,height=8.1cm]{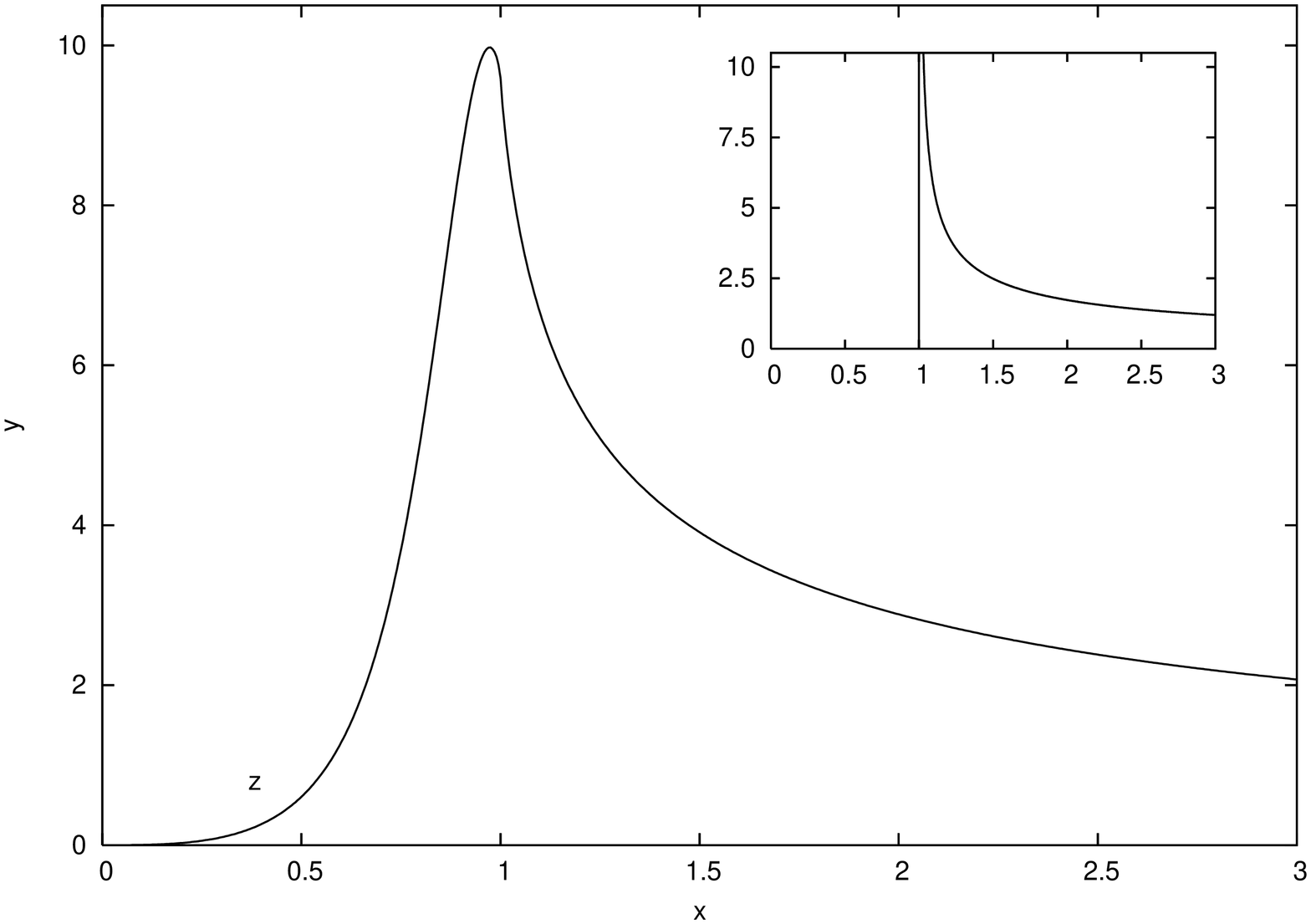}}
\caption{\label{fig:root} Left panel: the real (solid line) and imaginary (dashed line) parts of $\omega^2F_2-4\Delta^2F_4$ at $\xi=0$ and $T=0$. Right panel: the RPA Raman spectrum of an UDW for the chain-polarized scattering geometry, $x-x$ at $T=0$. Inset: the same RPA spectrum in a conventional DW.}
\end{figure}

In the small momentumtransfer limit ($\mathbf{q}\to0$), using the spinor introduced in Eq.~\eqref{spinor}, the interaction operator corresponding to the matrix element in Eq.~\eqref{int-matrix} can be recast as
\begin{equation}\label{int}
H_{\text{int}}=-\frac{P_i}{4}\Psi^\dag(\mathbf{k+q})A_i(\mathbf{k})\Psi(\mathbf{k})
\Psi^\dag(\mathbf{k'-q})A_i(\mathbf{k'})\Psi(\mathbf{k'}),
\end{equation}
where $i=c,s$ for UCDW and USDW, respectively. $A_c=\rho_1\sin(bk_y)$, $A_s=\rho_1\sigma_3\sin(bk_y)$ and the detailed form of the couplings $P_c$ for unconventional charge- and $P_s$ for spin-density waves are given in Ref.~[\onlinecite{balazs1}].
With this, the correlator of the "effective density" in the RPA reads as
\begin{subequations}\label{rpa-eq}
\begin{eqnarray}
\chi_{\gamma\gamma} &=& \chi^0_{\gamma\gamma}+\frac{P_iV_c}{4}\chi^0_{\gamma A_i}\chi_{A_i\gamma},\\
\chi_{A_i\gamma} &=& \chi^0_{A_i\gamma}+\frac{P_iV_c}{4}\chi^0_{A_iA_i}\chi_{A_i\gamma},
\end{eqnarray}
\end{subequations}
where $V_c$ is the cell volume and now the zero superscript denotes the one-bubble result. $\chi^0_{\gamma\gamma}$ is already given in Eq.~\eqref{raman}. In addition we obtain
\begin{subequations}
\begin{eqnarray}
\chi^0_{A_iA_i}(\xi,\omega) &=& 2g(0)\left(\frac{2}{\rho(0)P_i}+(\omega^2-\xi^2)F_2-4\Delta^2F_4\right),\label{int-int}\\
\chi^0_{\gamma A_i}(\xi,\omega) &=& 2g(0)\gamma_x\Delta\left(\frac{2}{\rho(0)P_i}+\omega^2F_2-\frac{4\Delta^2\omega^2}{\omega^2-\xi^2}F_4\right).\label{cross}
\end{eqnarray}
\end{subequations}
It is readily seen from Eq.~\eqref{cross} that the contributions from the $y-y$ and $z-z$ scattering geometries do not appear in the nondiagonal $\chi^0_{\gamma A_i}$ susceptibility, allowing only one Dyson series to develop, the one which dresses the chain polarized response, $x-x$. That is, taking into account the collective degrees of freedom of the DW, namely the fluctuation of the phase and amplitude of the order parameter around its mean field value, only this latter spectrum gets renormalized, while the former ones with polarizations aligned perpendicular to the quasi-one-dimensional direction retain their single-particle form. This situation is similar to the DC or optical conductivity of density waves, where the phase mode of the condensate contributes only to the chain direction as well [\onlinecite{balazs3}]. Solving the coupled RPA equations in Eq.~\eqref{rpa-eq}, for the full susceptibility we have
\begin{equation}\label{fullresponse}
\chi_{\gamma\gamma} = \chi^0_{\gamma\gamma}+\frac{P_iV_c}{4}\frac{\left(\chi^0_{\gamma A_i}\right)^2}{1-\frac{P_iV_c}{4}\chi_{A_iA_i}}.
\end{equation}
From this the Raman intensity is obtained as
\begin{equation}
\begin{split}
\text{Im}\chi^x_{\gamma\gamma} &= \text{Im}\frac{\chi^{0,x}_{\gamma\gamma}}{1-\frac{P_iV_c}{4}\chi^0_{A_iA_i}},\\
&=8g(0)\gamma_x^2(\Delta/\lambda)^2\frac{\omega^2F_2''-4\Delta^2F_4''}{|\omega^2F_2-4\Delta^2F_4|^2},
\end{split}
\end{equation}
where $\lambda=g(0)P_iV_c$ is the dimensionless coupling. The plot is shown in Fig.~\ref{fig:root} at $T=0$. The nature of the collective mode can be simply explored as usual by looking at the poles of the response function in Eq.~\eqref{fullresponse}. With Eq.~\eqref{int-int} the task is reduced to finding the roots of
\begin{equation}\label{root}
(\omega^2-\xi^2)F_2-4\Delta^2F_4=0.
\end{equation}

We note here that a similar expression ($(\omega^2-\xi^2)F_2=0$) appears when considering the density-density correlator of an (U)DW [\onlinecite{balazs3}], from which one obtains - for both conventional and unconventional DWs - the well known phason dispersion $\omega^2=\xi^2$ [\onlinecite{lra}]. In our case for a conventional DW, since $F_2=F_4$ (in Eq.~\eqref{f-function} $\sin(y)$ has to be replaced by 1, because neither the interaction nor the order parameter depends on the wavenumber) we reobtain the gapped dispersion of the amplitude excitation $\omega^2=4\Delta^2+\xi^2$ [\onlinecite{lra}], which is known to couple to the Raman experiment [\onlinecite{tsang,travaglini,gruner-book}].

Now turning our attention to the unconventional situation and remembering the result of the previous case, we look for the root on the real frequency axis around $2\Delta$. Therefore - after analytic continuation of the $F_n$'s for real $\omega$ - we plot the lhs. of Eq.~\eqref{root} versus frequency in the $\xi\to0$ limit, see Fig.~\ref{fig:root}, left panel. It is clear that - unlike what is found in Ref.~[\onlinecite{sumanta}] for a $d$-density-wave - there is no zero in either the real or the imaginary part at $2\Delta$, or at any other real frequency. There is no indication, that there would be a root of Eq.~\eqref{root} for complex frequency either. It follows that in contrast to conventional systems, in UDWs although the Raman intensity shows considerable renormalization due to electron-electron interaction with respect to the one-particle form, there is no clear, particle-like mode with infinite lifetime with which we could identify the peak around $\omega=2\Delta$, see Fig.~\ref{fig:root}, right panel. We can say however that the Raman vertex couples to the amplitude mode of the condensate, overdamped because of the low energy excitations.

\subsection{Coulomb screening}

In the case when the light produces charge fluctuation in the electron gas, the coupling to the long-range Coulomb forces reduces the scattering rate. Therefore it is useful to treat these forces separately. In the usual RPA approach the screened Raman susceptibility reads as
\begin{equation}\label{screened}
\chi_{\gamma\gamma}^{\text{sc}}=\chi_{\gamma\gamma}-\chi_{\gamma 1}\left(
V-V\chi_{11}V+\ldots\right)\chi_{1\gamma}=\chi_{\gamma\gamma}-\frac{\chi_{\gamma1}\chi_{1\gamma}}{\chi_{11}}
+\frac{\chi_{\gamma1}\chi_{1\gamma}}{\chi_{11}^2}\chi_{11}^{\text{sc}},
\end{equation}
where $V=4\pi e^2/q^2$, $\chi_{\gamma\gamma}$ is already calculated in Eq.~\eqref{raman}, $\chi_{11}^{\text{sc}}=\chi_{11}/(1+V\chi_{11})$ is the screened density correlator with $\chi_{11}$ being the single-particle contribution in unconventional charge- and spin-density waves given in Ref.~[\onlinecite{balazs3}]. The pole of the screened density correlator leads to the plasmon mode [\onlinecite{wang}], which however does not affect the low energy behavior. Furthermore, the nondiagonal $\chi_{\gamma1}$ term is expressed as
\begin{equation}\label{cross-coulomb}
\chi_{\gamma 1}=-g(0)\gamma_x(2\mu+\omega)\frac{\xi^2}{\xi^2-\omega^2}(1-4\Delta^2F_2).
\end{equation}
At this point it is important to call the attention to the fact, that similarly as in the previous subsection dealing with the short wavelength component of the interaction, here again we see that only one component of the whole Raman vertex survives in Eq.~\eqref{cross-coulomb}, namely the $x-x$ component. It means that in principle only the Raman spectrum with the incoming and scattered polarizations aligned in the chain direction can be screened by the Coulomb forces. Nevertheless, since both the $\chi_{11}$ and $\chi_{\gamma 1}$ correlators are quadratic in momentum in the long wavelength limit, thus the coupling between these quantities is not that strong to modify the zeroth order term $\chi_{\gamma\gamma}$ in Eq.~\eqref{screened}. The above calculations therefore lead us to the conclusion that the Raman response is not affected even if we take into account Coulomb screening in RPA. This is due to the vanishing average of the Raman vertex on the Fermi surface.

\section{Conclusions}

We have investigated theoretically the electronic Raman scattering in quasi-one dimensional interacting electron systems with density wave ground state. Mean field treatment of conventional as well as unconventional density waves in pure systems has been applied in order to determine the Raman intensity in various scattering geometries. We have found distinct, characteristic lineshapes especially in the unconventional situation, depending on the particular momentum dependence of the density wave order parameter. We conclude, that the Raman experiment could serve as a valuable tool in identifying materials supporting unconventional density waves, and in specifying their particular gap structure. We have also considered Coulomb screening, and we found it ineffective due to the negligible coupling of density fluctuations to the Raman vertex in our nearest neighbor tight-binding model. Collective contributions to the Raman response appear only in the $x-x$ scattering geometry (light polarized in the chain direction). This is due to coupling to the amplitude mode of the condensate. This mode is overdamped in the unconventional case, since decay to low energy excitations is possible.

\begin{acknowledgments}
We have benefited from discussions with R. Hackl and K. Maki. Special thanks go to B. D\'ora for his useful and instructive advices. This work was supported by the Hungarian Scientific Research Fund under Grants No. OTKA T046269, NDF45172 and TS040878.
\end{acknowledgments}

\bibliography{raman}

\begin{thebibliography}{38}
\expandafter\ifx\csname natexlab\endcsname\relax\def\natexlab#1{#1}\fi
\expandafter\ifx\csname bibnamefont\endcsname\relax
  \def\bibnamefont#1{#1}\fi
\expandafter\ifx\csname bibfnamefont\endcsname\relax
  \def\bibfnamefont#1{#1}\fi
\expandafter\ifx\csname citenamefont\endcsname\relax
  \def\citenamefont#1{#1}\fi
\expandafter\ifx\csname url\endcsname\relax
  \def\url#1{\texttt{#1}}\fi
\expandafter\ifx\csname urlprefix\endcsname\relax\def\urlprefix{URL }\fi
\providecommand{\bibinfo}[2]{#2}
\providecommand{\eprint}[2][]{\url{#2}}

\bibitem[{\citenamefont{Tsang et~al.}(1976)\citenamefont{Tsang, J.~E.~Smith,
  and Shafer}}]{tsang}
\bibinfo{author}{\bibfnamefont{J.~C.} \bibnamefont{Tsang}},
  \bibinfo{author}{\bibfnamefont{J.}~\bibnamefont{J.~E.~Smith}},
  \bibnamefont{and} \bibinfo{author}{\bibfnamefont{M.~W.}
  \bibnamefont{Shafer}}, \bibinfo{journal}{Phys. Rev. Lett.}
  \textbf{\bibinfo{volume}{37}}, \bibinfo{pages}{1407} (\bibinfo{year}{1976}).

\bibitem[{\citenamefont{Sooryakumar and Klein}(1980)}]{sooryakumar}
\bibinfo{author}{\bibfnamefont{R.}~\bibnamefont{Sooryakumar}} \bibnamefont{and}
  \bibinfo{author}{\bibfnamefont{M.~V.} \bibnamefont{Klein}},
  \bibinfo{journal}{Phys. Rev. Lett.} \textbf{\bibinfo{volume}{45}},
  \bibinfo{pages}{660} (\bibinfo{year}{1980}).

\bibitem[{\citenamefont{Behera and Bhattacharya}(1989)}]{Behera1}
\bibinfo{author}{\bibfnamefont{S.~N.} \bibnamefont{Behera}} \bibnamefont{and}
  \bibinfo{author}{\bibfnamefont{S.}~\bibnamefont{Bhattacharya}},
  \bibinfo{journal}{Phase Transitions} \textbf{\bibinfo{volume}{19}},
  \bibinfo{pages}{15} (\bibinfo{year}{1989}).

\bibitem[{\citenamefont{Behera and Bhattacharya}(1990)}]{Behera2}
\bibinfo{author}{\bibfnamefont{S.~N.} \bibnamefont{Behera}} \bibnamefont{and}
  \bibinfo{author}{\bibfnamefont{S.}~\bibnamefont{Bhattacharya}},
  \bibinfo{journal}{Physica C} \textbf{\bibinfo{volume}{167}},
  \bibinfo{pages}{112} (\bibinfo{year}{1990}).

\bibitem[{\citenamefont{Behera and Ghosh}(1994)}]{Behera3}
\bibinfo{author}{\bibfnamefont{S.~N.} \bibnamefont{Behera}} \bibnamefont{and}
  \bibinfo{author}{\bibfnamefont{H.}~\bibnamefont{Ghosh}}, \bibinfo{journal}{Z.
  Phys. B.} \textbf{\bibinfo{volume}{95}}, \bibinfo{pages}{275}
  (\bibinfo{year}{1994}).

\bibitem[{\citenamefont{Ghosh and Sardar}(1997)}]{Behera4}
\bibinfo{author}{\bibfnamefont{H.}~\bibnamefont{Ghosh}} \bibnamefont{and}
  \bibinfo{author}{\bibfnamefont{M.}~\bibnamefont{Sardar}},
  \bibinfo{journal}{Physica C} \textbf{\bibinfo{volume}{288}},
  \bibinfo{pages}{121} (\bibinfo{year}{1997}).

\bibitem[{\citenamefont{Devereaux et~al.}(1994)\citenamefont{Devereaux, Einzel,
  Stadlober, Hackl, Leach, and Neumeier}}]{leach}
\bibinfo{author}{\bibfnamefont{T.~P.} \bibnamefont{Devereaux}},
  \bibinfo{author}{\bibfnamefont{D.}~\bibnamefont{Einzel}},
  \bibinfo{author}{\bibfnamefont{B.}~\bibnamefont{Stadlober}},
  \bibinfo{author}{\bibfnamefont{R.}~\bibnamefont{Hackl}},
  \bibinfo{author}{\bibfnamefont{D.~H.} \bibnamefont{Leach}}, \bibnamefont{and}
  \bibinfo{author}{\bibfnamefont{J.~J.} \bibnamefont{Neumeier}},
  \bibinfo{journal}{Phys. Rev. Lett.} \textbf{\bibinfo{volume}{72}},
  \bibinfo{pages}{396} (\bibinfo{year}{1994}).

\bibitem[{\citenamefont{Snow et~al.}(2003)\citenamefont{Snow, Karpus, Cooper,
  Kidd, and Chiang}}]{snow}
\bibinfo{author}{\bibfnamefont{C.~S.} \bibnamefont{Snow}},
  \bibinfo{author}{\bibfnamefont{J.~F.} \bibnamefont{Karpus}},
  \bibinfo{author}{\bibfnamefont{S.~L.} \bibnamefont{Cooper}},
  \bibinfo{author}{\bibfnamefont{T.~E.} \bibnamefont{Kidd}}, \bibnamefont{and}
  \bibinfo{author}{\bibfnamefont{T.-C.} \bibnamefont{Chiang}},
  \bibinfo{journal}{Phys. Rev. Lett.} \textbf{\bibinfo{volume}{91}},
  \bibinfo{pages}{136402} (\bibinfo{year}{2003}).

\bibitem[{\citenamefont{Benfatto et~al.}(2000)\citenamefont{Benfatto, Caprara,
  and Castro}}]{benfatto}
\bibinfo{author}{\bibfnamefont{L.}~\bibnamefont{Benfatto}},
  \bibinfo{author}{\bibfnamefont{S.}~\bibnamefont{Caprara}}, \bibnamefont{and}
  \bibinfo{author}{\bibfnamefont{C.~D.} \bibnamefont{Castro}},
  \bibinfo{journal}{Eur. Phys. J. B} \textbf{\bibinfo{volume}{17}},
  \bibinfo{pages}{95} (\bibinfo{year}{2000}).

\bibitem[{\citenamefont{Bena et~al.}(2004)\citenamefont{Bena, Chakravarty, Hu,
  and Nayak}}]{bena}
\bibinfo{author}{\bibfnamefont{C.}~\bibnamefont{Bena}},
  \bibinfo{author}{\bibfnamefont{S.}~\bibnamefont{Chakravarty}},
  \bibinfo{author}{\bibfnamefont{J.}~\bibnamefont{Hu}}, \bibnamefont{and}
  \bibinfo{author}{\bibfnamefont{C.}~\bibnamefont{Nayak}},
  \bibinfo{journal}{Phys. Rev. B} \textbf{\bibinfo{volume}{69}},
  \bibinfo{pages}{134517} (\bibinfo{year}{2004}).

\bibitem[{\citenamefont{Chakravarty et~al.}(2001)\citenamefont{Chakravarty,
  Laughlin, Morr, and Nayak}}]{nayak}
\bibinfo{author}{\bibfnamefont{S.}~\bibnamefont{Chakravarty}},
  \bibinfo{author}{\bibfnamefont{R.~B.} \bibnamefont{Laughlin}},
  \bibinfo{author}{\bibfnamefont{D.~K.} \bibnamefont{Morr}}, \bibnamefont{and}
  \bibinfo{author}{\bibfnamefont{C.}~\bibnamefont{Nayak}},
  \bibinfo{journal}{Phys. Rev. B} \textbf{\bibinfo{volume}{63}},
  \bibinfo{pages}{094503} (\bibinfo{year}{2001}).

\bibitem[{\citenamefont{Zeyher and Greco}(2002)}]{zeyher}
\bibinfo{author}{\bibfnamefont{R.}~\bibnamefont{Zeyher}} \bibnamefont{and}
  \bibinfo{author}{\bibfnamefont{A.}~\bibnamefont{Greco}},
  \bibinfo{journal}{Phys. Rev. Lett.} \textbf{\bibinfo{volume}{89}},
  \bibinfo{pages}{177004} (\bibinfo{year}{2002}).

\bibitem[{\citenamefont{Castro et~al.}()\citenamefont{Castro, Grilli, Caprara,
  and Suppa}}]{castro}
\bibinfo{author}{\bibfnamefont{C.~D.} \bibnamefont{Castro}},
  \bibinfo{author}{\bibfnamefont{M.}~\bibnamefont{Grilli}},
  \bibinfo{author}{\bibfnamefont{S.}~\bibnamefont{Caprara}}, \bibnamefont{and}
  \bibinfo{author}{\bibfnamefont{D.}~\bibnamefont{Suppa}},
  \bibinfo{note}{cond-mat/0408058}.

\bibitem[{\citenamefont{Tacon et~al.}(2005)\citenamefont{Tacon, Sacuto, and
  Colson}}]{tacon}
\bibinfo{author}{\bibfnamefont{M.~L.} \bibnamefont{Tacon}},
  \bibinfo{author}{\bibfnamefont{A.}~\bibnamefont{Sacuto}}, \bibnamefont{and}
  \bibinfo{author}{\bibfnamefont{D.}~\bibnamefont{Colson}},
  \bibinfo{journal}{Phys. Rev. B} \textbf{\bibinfo{volume}{71}},
  \bibinfo{pages}{100504} (\bibinfo{year}{2005}).

\bibitem[{\citenamefont{Caprara et~al.}()\citenamefont{Caprara, Castro, Grilli,
  and Suppa}}]{caprara}
\bibinfo{author}{\bibfnamefont{S.}~\bibnamefont{Caprara}},
  \bibinfo{author}{\bibfnamefont{C.~D.} \bibnamefont{Castro}},
  \bibinfo{author}{\bibfnamefont{M.}~\bibnamefont{Grilli}}, \bibnamefont{and}
  \bibinfo{author}{\bibfnamefont{D.}~\bibnamefont{Suppa}},
  \bibinfo{note}{cond-mat/0501671}.

\bibitem[{\citenamefont{J\'anossy et~al.}(1997)\citenamefont{J\'anossy,
  Feh\'er, Oszl\'anyi, and Williams}}]{janossy}
\bibinfo{author}{\bibfnamefont{A.}~\bibnamefont{J\'anossy}},
  \bibinfo{author}{\bibfnamefont{T.}~\bibnamefont{Feh\'er}},
  \bibinfo{author}{\bibfnamefont{G.}~\bibnamefont{Oszl\'anyi}},
  \bibnamefont{and} \bibinfo{author}{\bibfnamefont{G.~V.~M.}
  \bibnamefont{Williams}}, \bibinfo{journal}{Phys. Rev. Lett.}
  \textbf{\bibinfo{volume}{79}}, \bibinfo{pages}{2726} (\bibinfo{year}{1997}).

\bibitem[{\citenamefont{Opel et~al.}(2000)\citenamefont{Opel, Nemetschek,
  Hoffmann, Philipp, M\"uller, Hackl, T\"utt{\H o}, Erb, Revaz, Walker
  et~al.}}]{opel}
\bibinfo{author}{\bibfnamefont{M.}~\bibnamefont{Opel}},
  \bibinfo{author}{\bibfnamefont{R.}~\bibnamefont{Nemetschek}},
  \bibinfo{author}{\bibfnamefont{C.}~\bibnamefont{Hoffmann}},
  \bibinfo{author}{\bibfnamefont{R.}~\bibnamefont{Philipp}},
  \bibinfo{author}{\bibfnamefont{P.~F.} \bibnamefont{M\"uller}},
  \bibinfo{author}{\bibfnamefont{R.}~\bibnamefont{Hackl}},
  \bibinfo{author}{\bibfnamefont{I.}~\bibnamefont{T\"utt{\H o}}},
  \bibinfo{author}{\bibfnamefont{A.}~\bibnamefont{Erb}},
  \bibinfo{author}{\bibfnamefont{B.}~\bibnamefont{Revaz}},
  \bibinfo{author}{\bibfnamefont{E.}~\bibnamefont{Walker}},
  \bibnamefont{et~al.}, \bibinfo{journal}{Phys. Rev. B}
  \textbf{\bibinfo{volume}{61}}, \bibinfo{pages}{9752} (\bibinfo{year}{2000}).

\bibitem[{\citenamefont{Kaminski et~al.}(2002)\citenamefont{Kaminski,
  Rosenkranz, Fretwell, Campuzano, Li, Raffy, Cullen, You, Olson, Varma
  et~al.}}]{kaminski}
\bibinfo{author}{\bibfnamefont{A.}~\bibnamefont{Kaminski}},
  \bibinfo{author}{\bibfnamefont{S.}~\bibnamefont{Rosenkranz}},
  \bibinfo{author}{\bibfnamefont{H.~M.} \bibnamefont{Fretwell}},
  \bibinfo{author}{\bibfnamefont{J.~C.} \bibnamefont{Campuzano}},
  \bibinfo{author}{\bibfnamefont{Z.}~\bibnamefont{Li}},
  \bibinfo{author}{\bibfnamefont{H.}~\bibnamefont{Raffy}},
  \bibinfo{author}{\bibfnamefont{W.~G.} \bibnamefont{Cullen}},
  \bibinfo{author}{\bibfnamefont{H.}~\bibnamefont{You}},
  \bibinfo{author}{\bibfnamefont{C.~G.} \bibnamefont{Olson}},
  \bibinfo{author}{\bibfnamefont{C.~M.} \bibnamefont{Varma}},
  \bibnamefont{et~al.}, \bibinfo{journal}{Nature}
  \textbf{\bibinfo{volume}{416}}, \bibinfo{pages}{610} (\bibinfo{year}{2002}).

\bibitem[{\citenamefont{N\'emeth et~al.}(2001)\citenamefont{N\'emeth, Matus,
  Kriza, and Alavi}}]{nemeth}
\bibinfo{author}{\bibfnamefont{L.}~\bibnamefont{N\'emeth}},
  \bibinfo{author}{\bibfnamefont{P.}~\bibnamefont{Matus}},
  \bibinfo{author}{\bibfnamefont{G.}~\bibnamefont{Kriza}}, \bibnamefont{and}
  \bibinfo{author}{\bibfnamefont{B.}~\bibnamefont{Alavi}},
  \bibinfo{journal}{Synth. Metals} \textbf{\bibinfo{volume}{120}},
  \bibinfo{pages}{1007} (\bibinfo{year}{2001}).

\bibitem[{\citenamefont{Isaacs et~al.}(1990)\citenamefont{Isaacs, McWhan,
  Kleiman, Bishop, Ice, Zschack, Gaulin, Mason, Garrett, and Buyers}}]{isaacs}
\bibinfo{author}{\bibfnamefont{E.~D.} \bibnamefont{Isaacs}},
  \bibinfo{author}{\bibfnamefont{D.~B.} \bibnamefont{McWhan}},
  \bibinfo{author}{\bibfnamefont{R.~N.} \bibnamefont{Kleiman}},
  \bibinfo{author}{\bibfnamefont{D.~J.} \bibnamefont{Bishop}},
  \bibinfo{author}{\bibfnamefont{G.~E.} \bibnamefont{Ice}},
  \bibinfo{author}{\bibfnamefont{P.}~\bibnamefont{Zschack}},
  \bibinfo{author}{\bibfnamefont{B.~D.} \bibnamefont{Gaulin}},
  \bibinfo{author}{\bibfnamefont{T.~E.} \bibnamefont{Mason}},
  \bibinfo{author}{\bibfnamefont{J.~D.} \bibnamefont{Garrett}},
  \bibnamefont{and} \bibinfo{author}{\bibfnamefont{W.~J.~L.}
  \bibnamefont{Buyers}}, \bibinfo{journal}{Phys. Rev. Lett.}
  \textbf{\bibinfo{volume}{65}}, \bibinfo{pages}{3185} (\bibinfo{year}{1990}).

\bibitem[{\citenamefont{Christ et~al.}(2000)\citenamefont{Christ, Biberacher,
  Kartsovnik, Steep, Balthes, Weiss, and M\"uller}}]{christ}
\bibinfo{author}{\bibfnamefont{P.}~\bibnamefont{Christ}},
  \bibinfo{author}{\bibfnamefont{W.}~\bibnamefont{Biberacher}},
  \bibinfo{author}{\bibfnamefont{M.~V.} \bibnamefont{Kartsovnik}},
  \bibinfo{author}{\bibfnamefont{E.}~\bibnamefont{Steep}},
  \bibinfo{author}{\bibfnamefont{E.}~\bibnamefont{Balthes}},
  \bibinfo{author}{\bibfnamefont{H.}~\bibnamefont{Weiss}}, \bibnamefont{and}
  \bibinfo{author}{\bibfnamefont{H.}~\bibnamefont{M\"uller}},
  \bibinfo{journal}{JETP Lett.} \textbf{\bibinfo{volume}{71}},
  \bibinfo{pages}{303} (\bibinfo{year}{2000}).

\bibitem[{\citenamefont{D\'ora and Virosztek}(2001)}]{balazs1}
\bibinfo{author}{\bibfnamefont{B.}~\bibnamefont{D\'ora}} \bibnamefont{and}
  \bibinfo{author}{\bibfnamefont{A.}~\bibnamefont{Virosztek}},
  \bibinfo{journal}{Eur. Phys. J. B} \textbf{\bibinfo{volume}{22}},
  \bibinfo{pages}{167} (\bibinfo{year}{2001}).

\bibitem[{\citenamefont{Ikeda and Ohashi}(1998)}]{ikeda}
\bibinfo{author}{\bibfnamefont{H.}~\bibnamefont{Ikeda}} \bibnamefont{and}
  \bibinfo{author}{\bibfnamefont{Y.}~\bibnamefont{Ohashi}},
  \bibinfo{journal}{Phys. Rev. Lett.} \textbf{\bibinfo{volume}{81}},
  \bibinfo{pages}{3723} (\bibinfo{year}{1998}).

\bibitem[{\citenamefont{Maki et~al.}(2003)\citenamefont{Maki, D\'ora,
  Kartsovnik, Virosztek, Korin-Hamzi\'c, and Basleti\'c}}]{balazs2}
\bibinfo{author}{\bibfnamefont{K.}~\bibnamefont{Maki}},
  \bibinfo{author}{\bibfnamefont{B.}~\bibnamefont{D\'ora}},
  \bibinfo{author}{\bibfnamefont{M.~V.} \bibnamefont{Kartsovnik}},
  \bibinfo{author}{\bibfnamefont{A.}~\bibnamefont{Virosztek}},
  \bibinfo{author}{\bibfnamefont{B.}~\bibnamefont{Korin-Hamzi\'c}},
  \bibnamefont{and}
  \bibinfo{author}{\bibfnamefont{M.}~\bibnamefont{Basleti\'c}},
  \bibinfo{journal}{Phys. Rev. Lett.} \textbf{\bibinfo{volume}{90}},
  \bibinfo{pages}{256402} (\bibinfo{year}{2003}).

\bibitem[{\citenamefont{D\'ora et~al.}(2003)\citenamefont{D\'ora, Maki,
  V\'anyolos, and Virosztek}}]{balazs4}
\bibinfo{author}{\bibfnamefont{B.}~\bibnamefont{D\'ora}},
  \bibinfo{author}{\bibfnamefont{K.}~\bibnamefont{Maki}},
  \bibinfo{author}{\bibfnamefont{A.}~\bibnamefont{V\'anyolos}},
  \bibnamefont{and}
  \bibinfo{author}{\bibfnamefont{A.}~\bibnamefont{Virosztek}},
  \bibinfo{journal}{Phys. Rev. B} \textbf{\bibinfo{volume}{68}},
  \bibinfo{pages}{241102(R)} (\bibinfo{year}{2003}).

\bibitem[{\citenamefont{D\'ora et~al.}(2004)\citenamefont{D\'ora, Maki,
  V\'anyolos, and Virosztek}}]{balazs5}
\bibinfo{author}{\bibfnamefont{B.}~\bibnamefont{D\'ora}},
  \bibinfo{author}{\bibfnamefont{K.}~\bibnamefont{Maki}},
  \bibinfo{author}{\bibfnamefont{A.}~\bibnamefont{V\'anyolos}},
  \bibnamefont{and}
  \bibinfo{author}{\bibfnamefont{A.}~\bibnamefont{Virosztek}},
  \bibinfo{journal}{Europhys. Lett.} \textbf{\bibinfo{volume}{67}},
  \bibinfo{pages}{1024} (\bibinfo{year}{2004}).

\bibitem[{\citenamefont{D\'ora et~al.}(2005)\citenamefont{D\'ora, Maki,
  Virosztek, and V\'anyolos}}]{balazs6}
\bibinfo{author}{\bibfnamefont{B.}~\bibnamefont{D\'ora}},
  \bibinfo{author}{\bibfnamefont{K.}~\bibnamefont{Maki}},
  \bibinfo{author}{\bibfnamefont{A.}~\bibnamefont{Virosztek}},
  \bibnamefont{and}
  \bibinfo{author}{\bibfnamefont{A.}~\bibnamefont{V\'anyolos}},
  \bibinfo{journal}{Phys. Rev. B} \textbf{\bibinfo{volume}{71}},
  \bibinfo{pages}{172502} (\bibinfo{year}{2005}).

\bibitem[{\citenamefont{Gavilano et~al.}(2004)\citenamefont{Gavilano, Rau,
  Pedrini, Hinderer, Ott, Kazakov, and Karpinski}}]{gavilano}
\bibinfo{author}{\bibfnamefont{J.~L.} \bibnamefont{Gavilano}},
  \bibinfo{author}{\bibfnamefont{D.}~\bibnamefont{Rau}},
  \bibinfo{author}{\bibfnamefont{B.}~\bibnamefont{Pedrini}},
  \bibinfo{author}{\bibfnamefont{J.}~\bibnamefont{Hinderer}},
  \bibinfo{author}{\bibfnamefont{H.~R.} \bibnamefont{Ott}},
  \bibinfo{author}{\bibfnamefont{S.~M.} \bibnamefont{Kazakov}},
  \bibnamefont{and}
  \bibinfo{author}{\bibfnamefont{J.}~\bibnamefont{Karpinski}},
  \bibinfo{journal}{Phys. Rev. B} \textbf{\bibinfo{volume}{69}},
  \bibinfo{pages}{100404(R)} (\bibinfo{year}{2004}).

\bibitem[{\citenamefont{Klein and Dierker}(1984)}]{dierker}
\bibinfo{author}{\bibfnamefont{M.~V.} \bibnamefont{Klein}} \bibnamefont{and}
  \bibinfo{author}{\bibfnamefont{S.~B.} \bibnamefont{Dierker}},
  \bibinfo{journal}{Phys. Rev. B} \textbf{\bibinfo{volume}{29}},
  \bibinfo{pages}{4976} (\bibinfo{year}{1984}).

\bibitem[{\citenamefont{Abrikosov and Genkin}(1973)}]{genkin}
\bibinfo{author}{\bibfnamefont{A.~A.} \bibnamefont{Abrikosov}}
  \bibnamefont{and} \bibinfo{author}{\bibfnamefont{V.~M.}
  \bibnamefont{Genkin}}, \bibinfo{journal}{Zh. Eksp. Teor. Fiz}
  \textbf{\bibinfo{volume}{65}}, \bibinfo{pages}{842} (\bibinfo{year}{1973}),
  \bibinfo{note}{[Sov. Phys. JETP \textbf{38}, 417 (1974)]}.

\bibitem[{\citenamefont{Virosztek and Maki}(1988)}]{viro2}
\bibinfo{author}{\bibfnamefont{A.}~\bibnamefont{Virosztek}} \bibnamefont{and}
  \bibinfo{author}{\bibfnamefont{K.}~\bibnamefont{Maki}},
  \bibinfo{journal}{Phys. Rev. B} \textbf{\bibinfo{volume}{37}},
  \bibinfo{pages}{2028} (\bibinfo{year}{1988}).

\bibitem[{\citenamefont{D\'ora and Virosztek}(2003)}]{balazs3}
\bibinfo{author}{\bibfnamefont{B.}~\bibnamefont{D\'ora}} \bibnamefont{and}
  \bibinfo{author}{\bibfnamefont{A.}~\bibnamefont{Virosztek}},
  \bibinfo{journal}{Europhys. Lett.} \textbf{\bibinfo{volume}{61}},
  \bibinfo{pages}{396} (\bibinfo{year}{2003}).

\bibitem[{\citenamefont{Devereaux and Einzel}(1995)}]{devereaux-1}
\bibinfo{author}{\bibfnamefont{T.~P.} \bibnamefont{Devereaux}}
  \bibnamefont{and} \bibinfo{author}{\bibfnamefont{D.}~\bibnamefont{Einzel}},
  \bibinfo{journal}{Phys. Rev. B} \textbf{\bibinfo{volume}{51}},
  \bibinfo{pages}{16336} (\bibinfo{year}{1995}).

\bibitem[{\citenamefont{Lee et~al.}(1974)\citenamefont{Lee, Rice, and
  Anderson}}]{lra}
\bibinfo{author}{\bibfnamefont{P.~A.} \bibnamefont{Lee}},
  \bibinfo{author}{\bibfnamefont{T.~M.} \bibnamefont{Rice}}, \bibnamefont{and}
  \bibinfo{author}{\bibfnamefont{P.~W.} \bibnamefont{Anderson}},
  \bibinfo{journal}{Solid State Commun.} \textbf{\bibinfo{volume}{14}},
  \bibinfo{pages}{703} (\bibinfo{year}{1974}).

\bibitem[{\citenamefont{Travaglini and Wachter}(1984)}]{travaglini}
\bibinfo{author}{\bibfnamefont{G.}~\bibnamefont{Travaglini}} \bibnamefont{and}
  \bibinfo{author}{\bibfnamefont{P.}~\bibnamefont{Wachter}},
  \bibinfo{journal}{Phys. Rev. B} \textbf{\bibinfo{volume}{30}},
  \bibinfo{pages}{1971} (\bibinfo{year}{1984}).

\bibitem[{\citenamefont{Gr\"uner}(1994)}]{gruner-book}
\bibinfo{author}{\bibfnamefont{G.}~\bibnamefont{Gr\"uner}},
  \emph{\bibinfo{title}{Density waves in solids}}
  (\bibinfo{publisher}{Addison-Wesley}, \bibinfo{address}{Reading},
  \bibinfo{year}{1994}).

\bibitem[{\citenamefont{Tewari and Chakravarty}(2002)}]{sumanta}
\bibinfo{author}{\bibfnamefont{S.}~\bibnamefont{Tewari}} \bibnamefont{and}
  \bibinfo{author}{\bibfnamefont{S.}~\bibnamefont{Chakravarty}},
  \bibinfo{journal}{Phys. Rev. B} \textbf{\bibinfo{volume}{66}},
  \bibinfo{pages}{054510} (\bibinfo{year}{2002}).

\bibitem[{\citenamefont{Wang et~al.}(2004)\citenamefont{Wang, Millis, and
  Sarma}}]{wang}
\bibinfo{author}{\bibfnamefont{D.~W.} \bibnamefont{Wang}},
  \bibinfo{author}{\bibfnamefont{A.~J.} \bibnamefont{Millis}},
  \bibnamefont{and} \bibinfo{author}{\bibfnamefont{S.~D.} \bibnamefont{Sarma}},
  \bibinfo{journal}{Solid State Commun.} \textbf{\bibinfo{volume}{131}},
  \bibinfo{pages}{637} (\bibinfo{year}{2004}).

\end{thebibliography}

\end{document}